# Machine Learning Harnesses Molecular Dynamics to Discover New μ Opioid Chemotypes

Evan N. Feinberg[Ω] | Amir Barati Farimani;[Ω] | Rajendra Uprety[Φ]| Amanda Hunkele[Φ] | Gavril W. Pasternak[Φ] | Susruta Majumdar[Φ] | Vijay S. Pande[Ψ,Ω]

Ψ → corresponding author. pande@stanford.edu
Ω → Stanford University
Φ → Sloan-Kettering Institute

ABSTRACT: Computational chemists typically assay drug candidates by virtually screening compounds against crystal structures of a protein despite the fact that some targets, like the μ Opioid Receptor and other members of the GPCR family, traverse many non-crystallographic states. We discover new conformational states of μOR with molecular dynamics simulation and then machine learn ligand-structure relationships to predict opioid ligand function. These artificial intelligence models identified a novel μ opioid chemotype.

## Introduction and Methodological Advance

A single class of proteins, the G-Protein Coupled Receptors (GPCRs), comprises over one-third of the targets of all FDA-approved drugs[1]. One such GPCR, the μ Opioid Receptor (μOR), epitomizes the benefits and drawbacks of existing GPCR drugs. Opioid chronic pain medications, such as morphine and hydrocodone, are μOR agonists that achieve their main therapeutic aim of analgesia, yet cause severe side effects, such as respiratory depression and addiction[2].

Like other GPCRs, μOR is not a binary switch. Rather, biophysical experiments indicate that GPCRs in general[3,4] and μOR in particular[5–9] traverse a spectrum of conformational states. *Our central thesis is that μOR samples a multiplicity of functionally relevant and pharmacologically predictive states. We have tested this hypothesis by developing novel computational methods to identify and incorporate these states, both yielding an increased AUC in opiate activity prediction and empowering the discovery of new opioid scaffolds.* Specifically, we posit that identifying important μOR states beyond the two crystal structures[10,11] would improve the ability to predict the activity of ligands at the Receptor.

To test this hypothesis, we conducted long timescale MD simulations of μOR either unliganded or bound to one of several agonists: BU72[12], Sufentanil[13], TRV130[14], and IBNtxA[7], providing a heterogeneous yet comprehensive spectrum of conformations that μOR can adopt. This dataset expands upon previous works[15,16] which focus on the conformational dynamics of the receptor. To systematically process such a large parallel MD dataset, consisting of an unprecedented 1.1 milliseconds of μOR simulation, we apply a kinetically motivated machine learning approach that (1) identified the reaction coordinates of μOR using the cutting-edge Sparse tICA algorithm[17–19] and (2) defined discrete receptor states using Minibatch K-Means clustering[20].

This *unsupervised* step uncovered key conformations of µOR, consisting both of intermediates between as well as non-canonical states distinct from the crystal structures (Figure 1, ref.[15]). In light of recent studies[21,22], these structures are potentially druggable states that can be directly employed to enrich rational drug discovery campaigns for µOR. To realize this potential, we trained *supervised* machine learning models that demonstrate significant improvement in two binary classifier tasks (Table 1): (1) the ability to distinguish agonists from antagonists, and (2) the ability to distinguish binders from non-binders at µOR.

Specifically, random forests[23], an increasingly prevalent machine learning tool in drug design, were deployed to connect structure to function. A database of opioids with known pharmacologies was docked to both the crystal structures as well as to a representative conformation of each state. The docking scores of each ligand to each MD conformation were then used as the input (E.D. Fig. 3), or feature matrix, to binary classifier[24] models for both agonism and binding at µOR.

**Statistical Results and Small Molecule Discovery**

This dually unsupervised and supervised ML-based synthesis of structural information from both crystallography and MD yielded statistically significant enrichment in both tasks. Model performance is evaluated using the Receiver Operating Characteristic (ROC) Area Under the Curve (AUC[24]) and stratified cross-validation (Methods). Cross-validation is a pseudo-prospective validation of the model which estimates the experimental performance on previously unseen query ligands. The new method – which incorporates docking to the MD states in addition to the crystal structures – achieved a median AUC improvement of 0.11 in agonism and of 0.15 in binding compared to the crystal structures alone (Table 1, E.D. Tables 4 and 5).

In contrast to many previous virtual screening approaches, which are predicated solely on ligand-derived features, the machine-learning approach described here is based on the affinities of a given ligand toward each µOR conformation. As a further test of this method's robustness to agonism, scaffold splits were employed. Specifically, a series of models were trained in which analogs of either methadone or fentanyl were removed from the training data and placed in a held-out test set (Table 1, E.D. Tables 2 and 4). In other words, none of these models had any *a priori* knowledge of methadone (or, alternately, fentanyl) analogs. Nevertheless, the models successfully distinguished both methadone- and fentanyl-derived agonists from random sets of antagonists. Analogous scaffold splits were defined for the binding prediction task, yielding comparable gains in AUC (Table 1, E.D. Table 5). Therefore, since this method does not explicitly incorporate the chemical makeup of the ligand, it is in principle better equipped to discover new opioid-active scaffolds in addition to derivatives of existing ones.

In practice this method identified several novel opioid-active ligands that were verified experimentally. First, 133,564 small molecules from the Stanford Compound Library were docked to both crystal structures and the computationally modeled conformers of µOR. This step yielded a 133,564 row by 27 column feature matrix, where entry (i, j) is the docking score of the *i*'th ligand to the *j*'th conformational state. The two trained random forest models for binding and agonism were applied to each library ligand, yielding a final score computed from the product of the two values:

$$P(binder \cap agonist \,|\, model) = P(binder \,|\, model_b) \cdot P(agonist \,|\, model_a)$$

Both model performance and scaffolds of hits are highly sensitive to the pIC50 cutoff chosen for binary classifiers (E.D. Table 5). While models with a lower affinity threshold for binding generally have higher AUC, their top hits are more biased toward compounds with a tertiary, basic nitrogen similar to known scaffolds. Therefore, we used random forests models with a pIC50 cutoff of 8.0 (10 nM) to optimize for novel scaffold discovery.

The thirty highest scoring, immediately purchasable compounds were then experimentally assayed. At least three of the thirty compounds exhibited micromolar affinity for µOR (E.D. Figure 4, E.D. Table 6). FMP4, because of its unique structure – namely, no basic amine or phenol – was characterized further in binding assays in opioid transfected cell lines. FMP4 had 3217±153 nM, 2503± 523 nM and 8143±1398 nM affinity at MOR-1, KOR-1 and DOR-1 respectively (E.D. Figure 5). FMP 4 was also a weak MOR-1 partial agonist in [$^{35}$S]GTPγS functional assays. (E.D. Figure 6). Structurally similar FMP4-like compounds in the same data set were also characterized in binding assays and at least two compounds (FMP1 and FMP16) showed <10µM affinity at MOR-1 (E.D. Table 7). Structure-activity studies using classical medicinal chemistry approaches using these templates as a starting point may lead to compounds with higher affinities at the receptor.

Based on these results, *opioid prediction is enriched by incorporating conformational states that are unforeseeable from crystallography alone and stabilized by ligands in simulation.* Notably, FMP4 is distinct from known opioid agonists and antagonists, with a maximum Tanimoto score of 0.44 to the ligands listed in E.D. Table 1. Underscoring its difference is FMP4's lack of a basic tertiary amine and phenol, that are the hallmark of synthetic opioids, from classic to current experimental molecules[5,7,14,25].

Modeling predicts that FMP4 binds to and facilitates activation of µOR in a unique way. In particular, FMP4 has a relatively high docking score for MD State 3, calculated to be important for agonism and binding (E.D. Table 3). Note that tIC.1, the slowest tICA reaction coordinate, connects the two crystallographic states (Figure 1). In contrast, tIC.2, the second slowest tICA coordinate, is kinetically orthogonal to tIC.1 and defines several non-crystallographic states (Figure 1b). Measured by its progress along tIC.1 and by traditional metrics of GPCR literature[26,27] such as outward orientation of transmembrane helix 6 and bulged configuration of the NPxxY motif residues N332$^{7.48}$-Y336$^{7.53}$, MD State 3 is a novel active-like state of µOR. Near the orthosteric binding site, State 3 entails a rearrangement of Q124$^{2.60}$, M151$^{3.36}$, H297$^{6.52}$, Y299$^{6.54}$, and W318$^{6.35}$. The new positions of M151$^{3.36}$ and H297$^{6.52}$ enable FMP4 to occupy a pose that would be sterically forbidden in the active crystal structure (Figure 1a). In contrast to the co-crystallized agonist, FMP4 engages in a π-T interaction with W293$^{6.48}$, a residue critical in gating µOR activation[10], and a hydrogen bond with H297$^{6.52}$. A PyMOL session and PDB files of each modeled µOR conformation is available in the Extended Data section, and further visualizations in E.D. Figure 1.

**Discussion**

Put together, *by enumerating the state space of µOR, one can query conformations of the receptor to motivate rational design with all-atom structural information*. An ineluctable flood of data stems from MD, and it is a significant data science challenge to derive actionable knowledge from a vast dataset of simulation alone. One millisecond of MD saved at one frame per nanosecond would contain one million conformations, far too many to be viewed by expert eyes. Rather, by pursuing a kinetically-motivated statistical approach, it was possible to discover key conformations of the receptor within a tractable scope.

A key element of this approach is the estimation of the affinity of each ligand for each of several conformations of a receptor. These conformations are obtained in a single MD simulation pre-step and span a structural basis set for the receptor's functionality. In contrast, induced fit docking[28–30] samples different conformations to estimate a single docking (affinity) score for the protein. However, its conformational sampling is spatiotemporally limited, extends only to the binding pocket, must be repeated for each ligand, and, by outputting a single number correlated with affinity it is intrinsically not targeted for predicting agonism.

The method described herein assists translational researchers in realizing a challenging goal: identifying novel GPCR drug scaffolds. Despite immense efforts in medicinal chemistry on synthesizing derivatives of existing chemotypes, all FDA-approved opioids are riddled with serious side effects that restrict their utility in treating acute and chronic pain. In this work, we describe a method to leverage crystallography and molecular modeling with machine learning to explore previously uncharted chemical space of molecules active at μOR. This approach is naturally applicable to any receptor which is expected to have any sort of conformational plasticity, including other GPCRs[31], kinases, ion channels, and nuclear receptors. The ultimate goal of superior opiate therapeutics remains in the future, but now may loom much closer on the horizon.

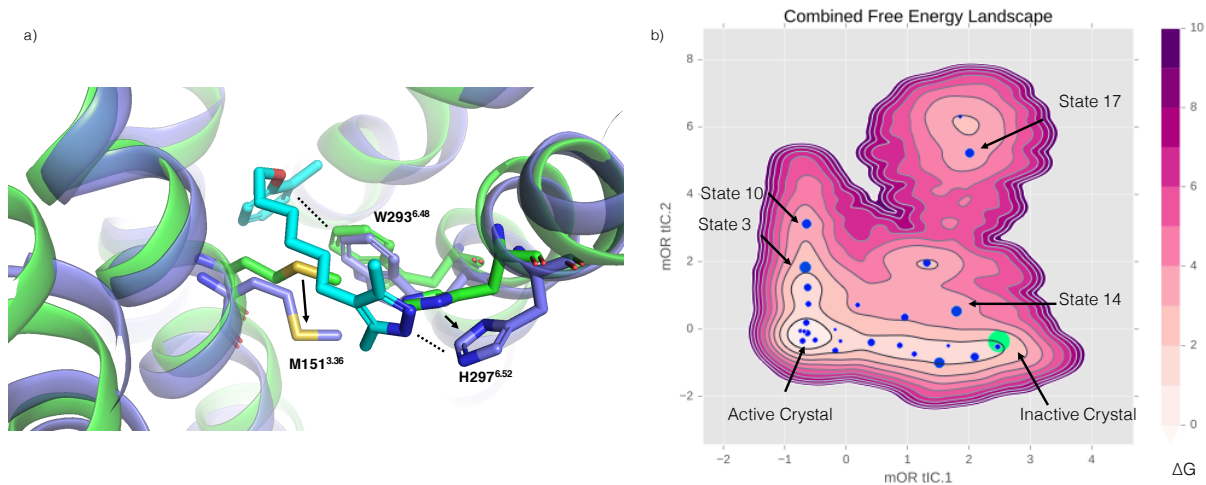

Figure 1: a) Active crystal structure (PDB: 5C1M) in green; MD State 3 in dark blue; docked pose of FMP4 to MD State 3 in cyan. Solid arrows represent changes in MD from crystal structures. Dashed lines indicate non-covalent interactions between FMP4 and μOR binding pocket residues. Note that FMP4 would sterically clash with residues M151 and H297 in the active crystal, possibly accounting for its very low docking score to that structure. Movements of M151 and H297 enable favorable noncovalent ligand-protein interactions in a non-strained conformation of the ligand. Unlike the morphinan phenol, FMP4's phenyl ring engages with key activation residue W293 with a π-T aromatic interaction. b) Free energy landscape of μOR projected onto its two slowest collective degrees of freedom. Whereas tICA coordinate 1 separates the active and inactive (PDB: 4DKL) crystal structures, tICA coordinate 2 is an orthogonal degree of freedom defining several non-crystallographic inactive and active-like states. Such states include State 3, critical for FMP4's ability to engage with the receptor.

| Task | Split Type | AUC (Crystals alone) | AUC (Crystal + MD structures) |
|---|---|---|---|
| Agonism | Random | 0.73 | **0.85** |
| Agonism | Scaffold (Fentanyl) | 0.81 | **0.91** |
| Agonism | Scaffold (Methadone) | 0.89 | **0.94** |
| Binding | Random | 0.64 | **0.79** |
| Binding | Scaffold | 0.64 | **0.78** |

Table 1: Docking to both MD and crystal structures statistically significantly improves ability over crystal structures alone to distinguish μOR agonists from antagonists and μOR binders from non-binders. Table shows median ROC Area Under the Curve (AUC) performance over 1,000 train-validation splits. When either fentanyl or methadone analogs are removed from the training set (scaffold splits, cf. Methods), models remain able to distinguish fentanyl (or methadone) derivative agonists from antagonists. Similarly, when ligands with a Tanimoto similarity score greater than 0.7 to any other ligand in the database are removed from the training data (scaffold split), AUC improves commensurately when building a model incorporating MD-derived conformers. Therefore, models fit in this way have, in principle, the capacity to discover new opioid scaffolds in addition to derivatives of existing ones.

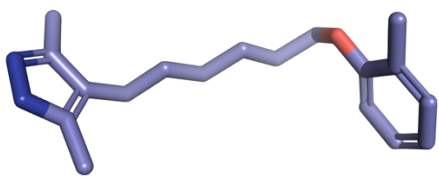

Scheme 1: Compound FMP4 (Pubchem ID: 2057658) is one of at least four purchased compounds with affinity for μOR at least at the 10 μM level. FMP4 specifically exhibits **3.2 μM** affinity with weak partial agonism as well.

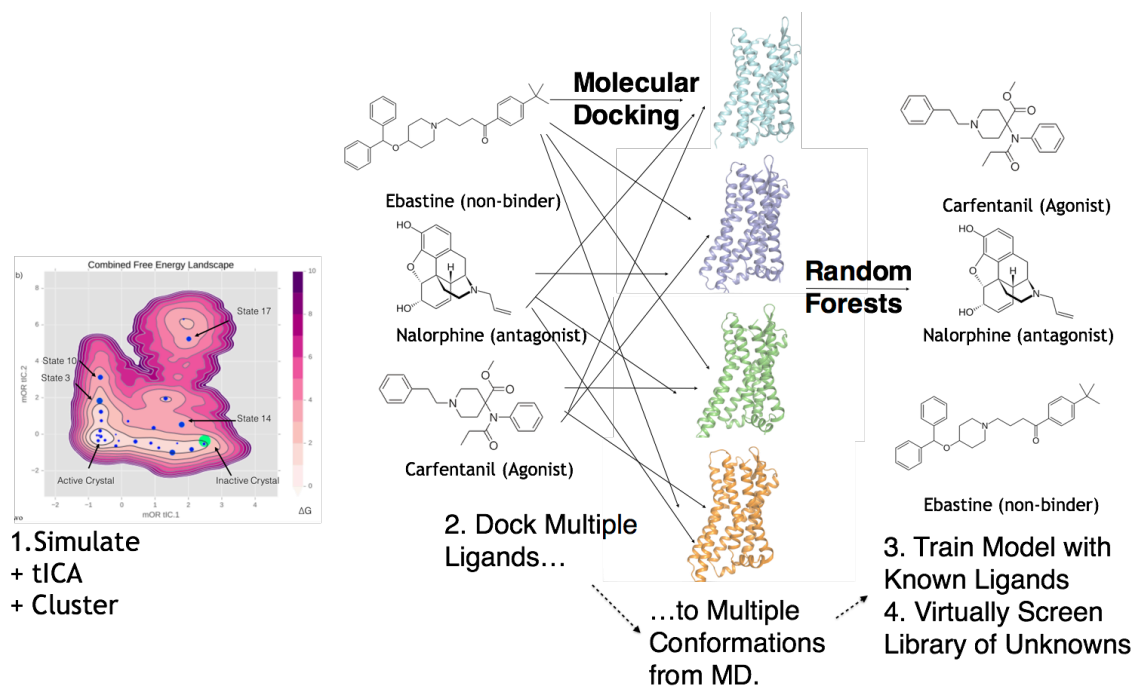

Figure 2: Method Overview

1. Overington, J.P., Al-Lazikani, B. & Hopkins, A.L. *Nat. Rev. Drug Discov.* **5**, 993–996 (2006).
2. Pasternak, G.W. & Pan, Y.-X. *Pharmacol. Rev.* **65**, 1257–1317 (2013).
3. Staus, D.P. et al. *Nature* **535**, 448–452 (2016).
4. Manglik, A. et al. *Cell* **161**, 1101–1111 (2015).
5. Manglik, A. et al. *Nature* 1–6 (2016).
6. Okude, J. et al. *Angew. Chemie* **127**, 15997–16002 (2015).
7. Majumdar, S. et al. *Proc. Natl. Acad. Sci.* **108**, 19778–19783 (2011).
8. Váradi, A. et al. *J. Med. Chem.* (2016).
9. Sounier, R. et al. *Nature* **524**, 375–378 (2015).
10. Huang, W. et al. *Nature* **524**, 315–21 (2015).
11. Manglik, A. et al. *Nature* **485**, 321–326 (2012).
12. Neilan, C.L. et al. *Eur. J. Pharmacol.* **499**, 107–116 (2004).
13. Niemegeers, C.J., Schellekens, K.H., Van Bever, W.F. & Janssen, P.A. *Arzneimittelforschung.* **26**, 1551–1556 (1975).
14. Chen, X.-T. et al. *J. Med. Chem.* **56**, 8019–8031 (2013).
15. Feinberg, E.N., Farimani, A.B., Hernandez, C.X. & Pande, V.S. *bioRxiv* 170886 (2017).
16. Huang, W. et al. *Nature* **524**, 315–21 (2015).
17. Schwantes, C.R. & Pande, V.S. *J. Chem. Theory Comput.* **9**, 2000–2009 (2013).
18. Pérez-Hernández, G., Paul, F., Giorgino, T., De Fabritiis, G. & Noé, F. *J. Chem. Phys.* **139**, 15102 (2013).
19. McGibbon, R.T., Husic, B.E. & Pande, V.S. *J. Chem. Phys.* **146**, 44109 (2017).
20. Sculley, D. *Proc. 19th Int. Conf. World wide web* 1177–1178 (2010).
21. Offutt, T.L., Swift, R. V & Amaro, R.E. *J. Chem. Inf. Model.* (2016).

22. Weiss, D.R. et al. *ACS Chem. Biol.* **8**, 1018–1026 (2013).
23. Breiman, L. *Mach. Learn.* **45**, 5–32 (2001).
24. Friedman, J., Hastie, T. & Tibshirani, R.**1**, (Springer series in statistics Springer, Berlin: 2001).
25. Goldberg, J.S. *Perspect. Medicin. Chem.* **4**, 1 (2010).
26. Dror, R.O. et al. *Proc. Natl. Acad. Sci.* **108**, 18684–18689 (2011).
27. Latorraca, N.R., Venkatakrishnan, A.J. & Dror, R.O. *Chem. Rev.* (2016).
28. Sherman, W., Beard, H.S. & Farid, R. *Chem. Biol. Drug Des.* **67**, 83–84 (2006).
29. Clark, A.J. et al. *J. Chem. Theory Comput.* **12**, 2990–2998 (2016).
30. Nabuurs, S.B., Wagener, M. & De Vlieg, J. *J. Med. Chem.* **50**, 6507–6518 (2007).
31. Feinberg, E.N. & Pande, V.S. *Prep* (2016).

**Acknowledgements**

The authors express their appreciation to Brooke E. Husic, Mohammad M. Sultan, Bharath Ramsundar, Nathaniel Stanley, Danielle J. Marshak, Steven Kearnes, and Samuel Hertig for their insightful input while crafting this manuscript. We also would like to thank Kilian Cavalotti, Brian Roberts, Jimmy Wu, and Stephane Thiell for their indispensable computing support. We thank Keri A. McKiernan for help debugging and improving the docking analysis code. We thank Susruta Majumdar at Sloan-Kettering in New York for refinement of the Wikipedia-based opioid labeling dataset. We thank NIH training grant T32 GM08294 and NIH grant 1171245-309-PADPO for funding, as well as acknowledge the use and the University of Illinois. This research was also funded, in part, by grants from the National Institute on Drug Abuse (DA0624), the Mayday Foundation and the Peter F. McManus Trust (G.W.P.), DA 034106 (S.M.) and an NIH/NCI Cancer Center Support Grant P30 CA008748 to Memorial Sloan Kettering Cancer Center.

**Author Contributions**

E.N.F. and V.S.P. conceived, and V.S.P. supervised, this project. A.B.F. and E.N.F. set up simulation systems. A.B.F. parameterized ligands in Gaussian for the CHARMM force field. E.N.F. generated the prospective docking poses of, and A.B.F. conducted (with NAMD) and analyzed the preliminary simulations of Sufentanil in various poses. E.N.F. conducted all other simulation (listed in E.D. Figure 7) in AMBER on XStream. E.N.F. wrote the Python-based code used in analysis of simulations. E.N.F. designed, trained, and tested the small molecule classifier models with consultation with S.M.. A.H., R.U. designed and conducted, the biochemical assays under S.M.'s supervision G.W.P. and S.M. edited the manuscript.

**Competing Financial Interests**

VSP is a consultant & SAB member of Schrodinger, LLC and Globavir, sits on the Board of Directors of Apeel Inc, Freenome Inc, Omada Health, Patient Ping, Rigetti Computing, and is a General Partner at Andreessen Horowitz.

# Machine Learning Harnesses Molecular Dynamics to Discover New μ Opioid Chemotypes

Evan N. Feinberg[Ω] | Amir Barati Farimani[Ω] | Rajendra Uprety[Φ] | Amanda Hunkele[Φ] | Gavril W. Pasternak[Φ] | Susruta Majumdar[Φ] | Vijay S. Pande[Ψ,Ω]

Ψ → corresponding author. pande@stanford.edu
Ω → Stanford University
Φ → Sloan-Kettering Institute

## Extended Data and Tables

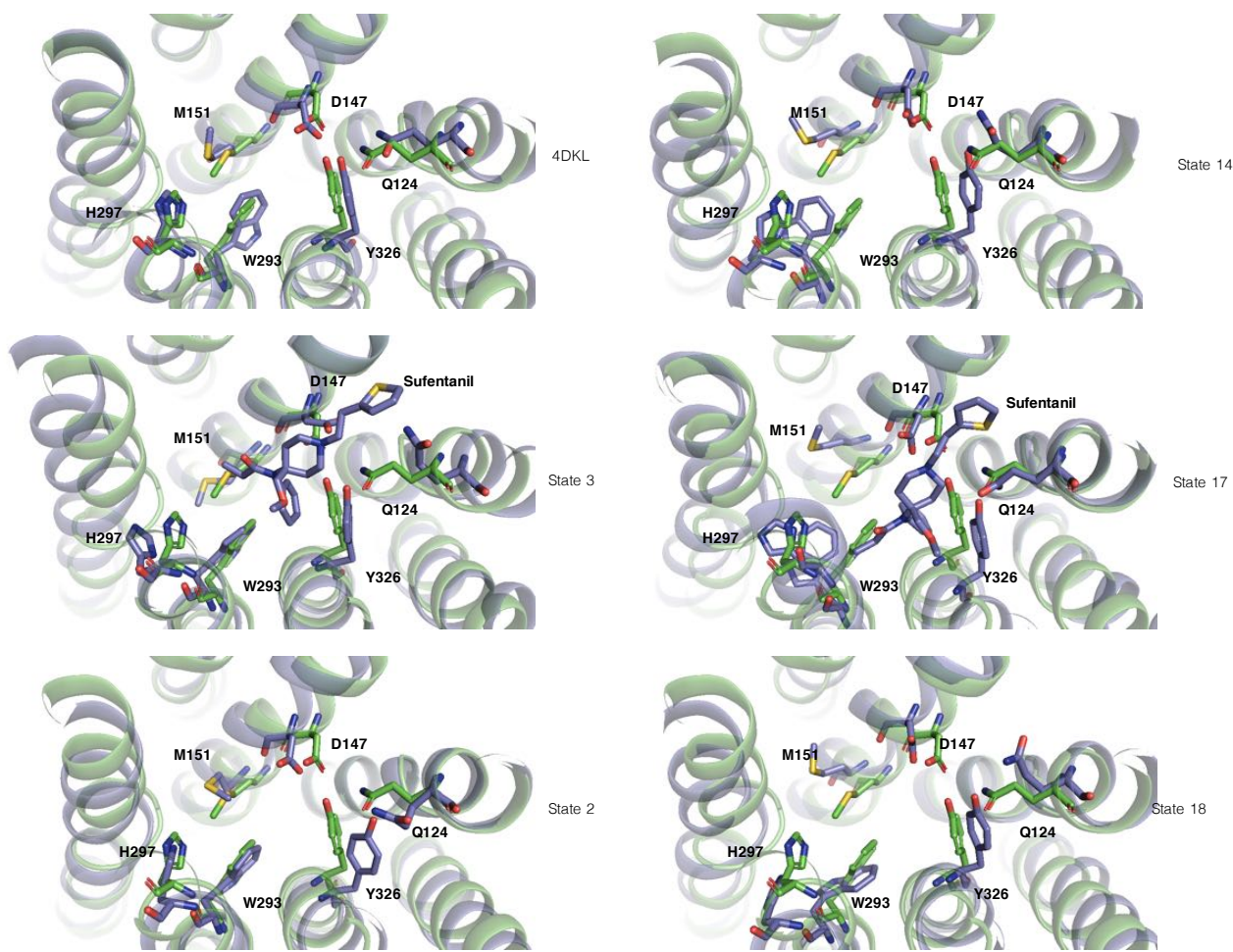

**Extended Data Figure 1**

Binding pocket views of the top three most important conformational states of μOR derived from MD simulation by measure of largest mean reduction in Gini impurity for random forest models of binding and of agonism. A PyMOL session as well as individual PDB files of each conformation can be downloaded in the supplemental information section.

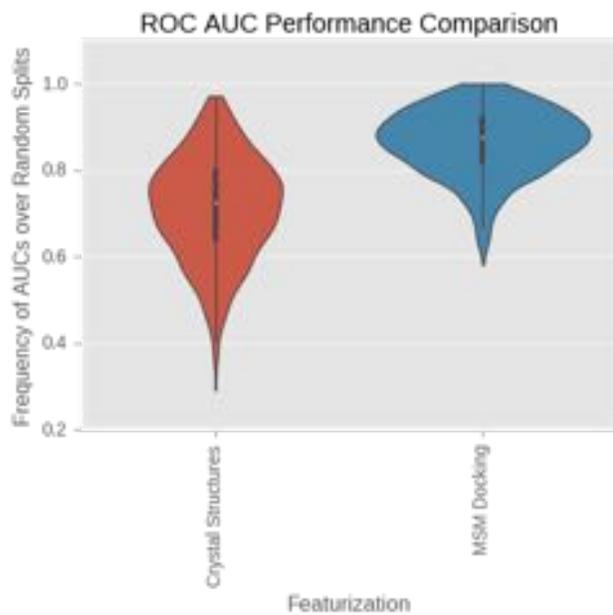

**Extended Data Figure 2**

Violin plots comparing the two methods' (docking to crystal structures alone versus docking to crystal structures and MSM states) AUC performance over 1000 random train-test splits.

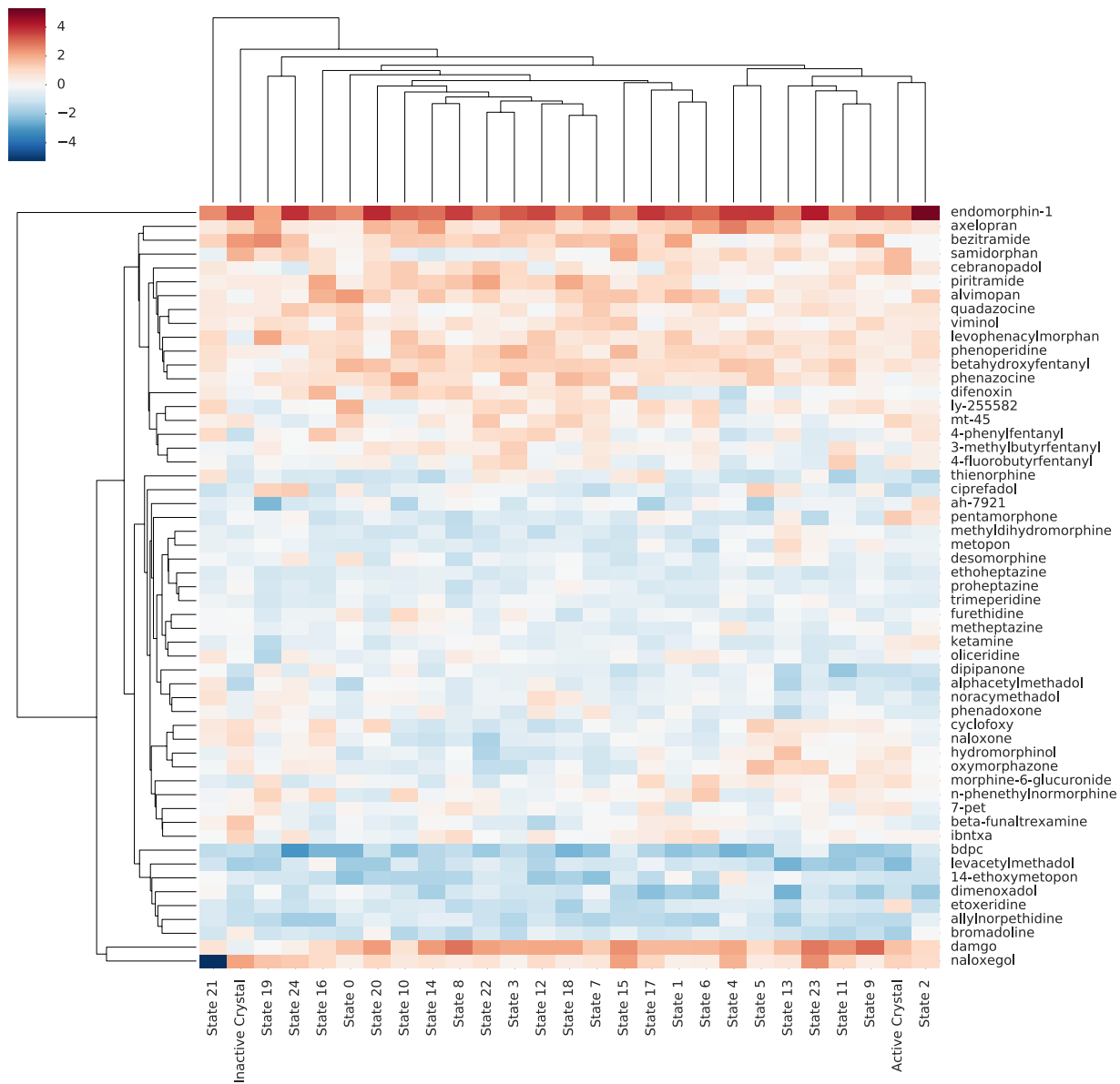

**Extended Data Figure 3**

Visualization of a subset of the feature matrix (X matrix), the input to the machine learning algorithm. Each row is an opioid ligand, each column is a feature (docking scores to each of MD States 0-24 and to each crystal structure). Therefore, entry (i, j) in the matrix is the docking score of the *i*'th ligand to the *j*'th conformational state. For visualization, each row is normalized to mean zero and unit variance. Also for visualization, Rows (ligands) and columns (features) are also hierarchically clustered.

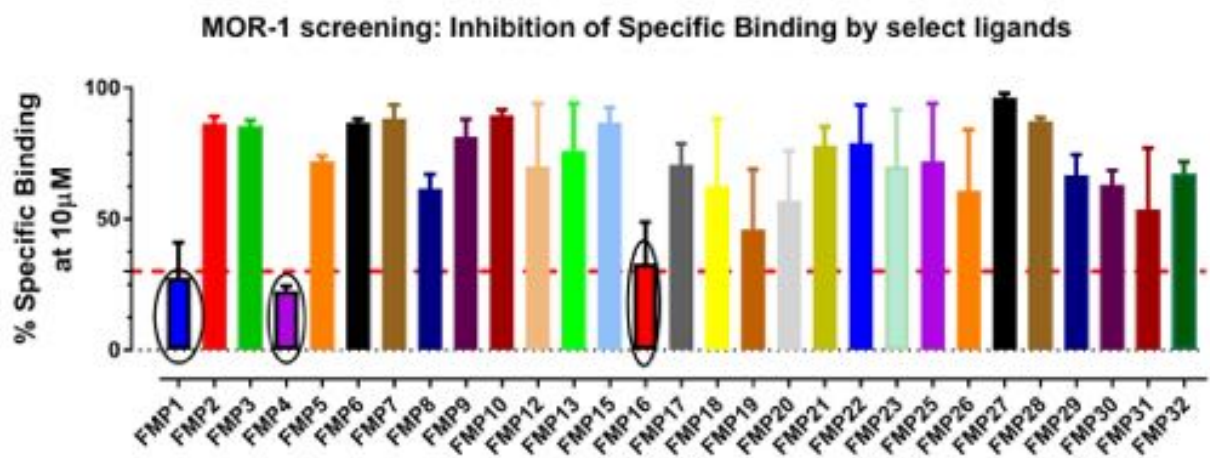

**Extended Data Figure 4. Screening of MOR-1 binders:** Inhibition of [125]IBNtxA specific binding at MOR-1 was carried out at a single dose 10 µM concentration. Three compounds FMP**1**, **4** and **16** (circled) showed ≥30% inhibition of MOR-1 binding (red dotted line represents compounds showing≥30% inhibition). Each panel is a representative experiment that has been independently replicated at least three times.

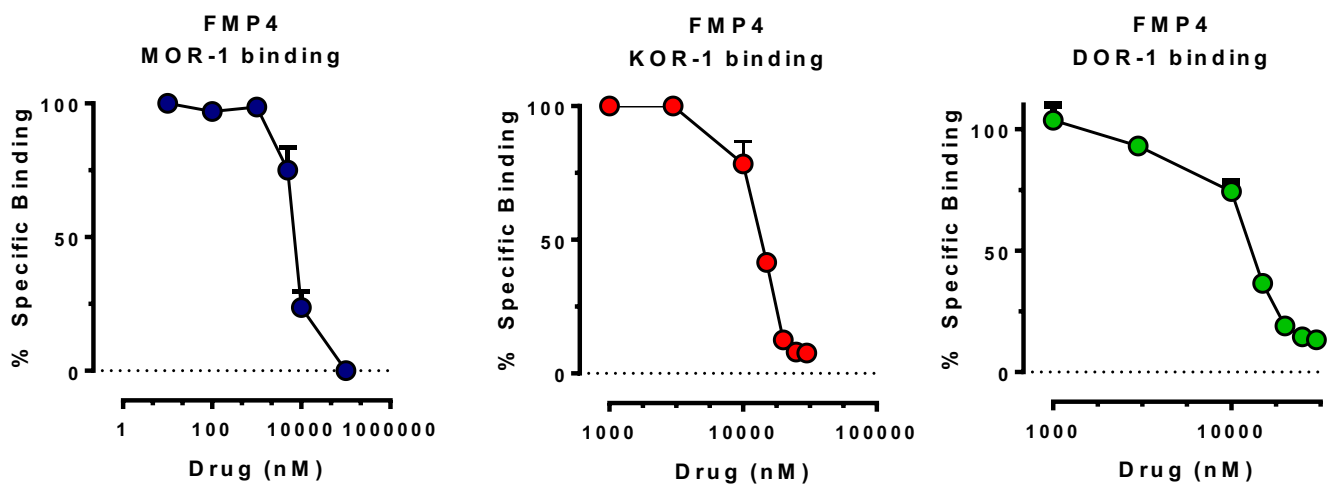

**Extended Data Figure 5. Competition assays at MOR-1, KOR-1 and DOR-1 against FMP4**. Competition studies were performed with FMP4 against $^{125}$I-IBNtxA (0.1 nM) in membranes from CHO cells stably expressing the indicated cloned mouse opioid receptors Each figure is a representative experiment that has been independently replicated at least three times. Error bars represent the SEM of triplicate samples. Error bars that cannot be seen are smaller than the size of the symbol. FMP4 had 3217±153 nM, 2503± 523 nM and 8143±1398 nM affinity at MOR-1, KOR-1 and DOR-1 respectively.

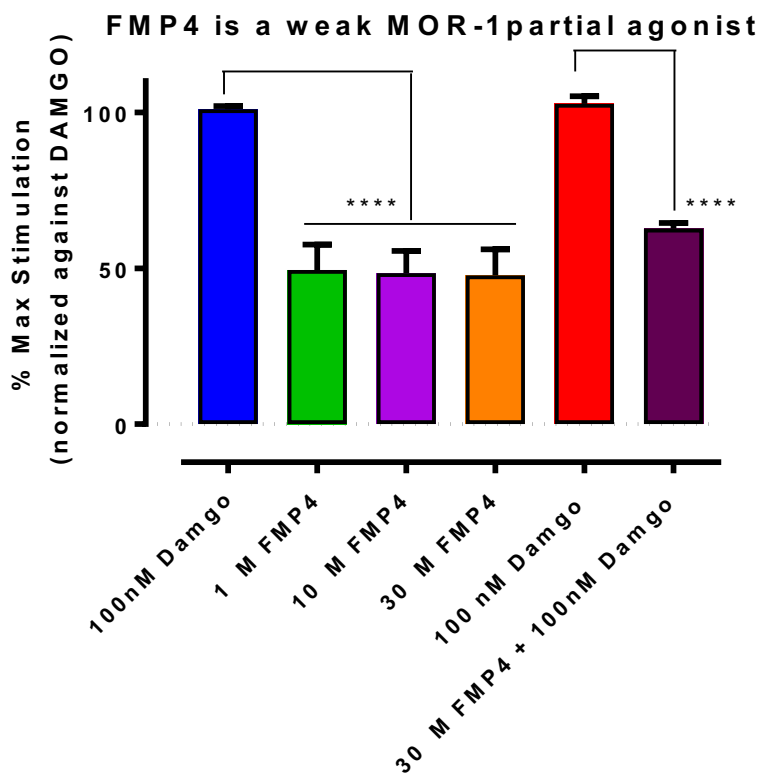

**Extended Figure Data 6: FMP4 is a partial agonist in [$^{35}$S]-GTPγS assays in MOR-1 expressing CHO cells.** Results are pooled from 3 independent replications and are expressed as mean ± SEM. **FMP4** stimulates GTPγS binding at doses 1, 10 and 30 µM respectively and shows a ceiling effect consistent with partial agonism at the receptor in comparison to the prototypic full MOR-1 agonist DAMGO at 100 nM (one-way ANOVA followed by Tukeys post hoc comparisons test, *significantly different from control ($p < 0.0001$)). **FMP4** at a dose of 30 µM also antagonized the GTPγS stimulation of 100 nM DAMGO consistent with its partial agonism at the receptor.

| Simulation Ligand | Aggregate MD Simulation, Generation 1 | Aggregate MD Simulation, Generation 2 | Total Number of Trajectories |
|---|---|---|---|
| Apo | 115.839 μs | 119.620 μs | 512 |
| BU72 | 102.705 μs | 145.794 μs | 512 |
| Sufentanil | 105.298 μs | 166.723 μs | 754 |
| IBNtxA | 93.491 μs | 0 μs | 256 |
| TRV130 | 241.360 μs | 0 μs | 512 |
|  | 1.090830 ms total |  | 2,546 |

**Extended Data Figure 7**

Summary of MD simulations conducted of μOR.

| name | action |
|---|---|
| 7-pet | agonist |
| acetylfentanyl | agonist |
| acetylmethadol | agonist |
| acrylfentanyl | agonist |
| ah-7921 | agonist |
| alfentanil | agonist |
| alimadol | agonist |
| 3-allylfentanyl | agonist |
| allylnorpethidine | agonist |
| allylprodine | agonist |
| alphacetylmethadol | agonist |
| alphamethadol | agonist |
| alphamethylthiofentanyl | agonist |
| anileridine | agonist |
| azaprocin | agonist |
| azidomorphine | agonist |
| bdpc | agonist |
| benzethidine | agonist |
| betacetylmethadol | agonist |
| betahydroxyfentanyl | agonist |
| betahydroxythiofentanyl | agonist |
| betamethadol | agonist |
| bezitramide | agonist |
| brifentanil | agonist |
| bromadoline | agonist |
| butyrfentanyl | agonist |
| c-8813 | agonist |
| carfentanil | agonist |
| cebranopadol | agonist |
| chloromorphide | agonist |
| chloroxymorphamine | agonist |
| ciprefadol | agonist |
| clonitazene | agonist |
| dadle | agonist |
| damgo | agonist |
| dermorphin | agonist |
| desmethylprodine | agonist |
| desomorphine | agonist |
| dextromoramide | agonist |
| dextropropoxyphene | agonist |
| diampromide | agonist |
| difenoxin | agonist |
| dihydroetorphine | agonist |
| dihydromorphine | agonist |
| dimenoxadol | agonist |
| dimepheptanol | agonist |
| dimethylaminopivalophenone | agonist |
| dioxaphetyl_butyrate | agonist |
| diphenoxylate | agonist |
| dipipanone | agonist |
| dpi-3290 | agonist |
| eluxadoline | agonist |
| endomorphin | agonist |
| endomorphin-1 | agonist |
| endomorphin-2 | agonist |
| ethoheptazine | agonist |
| 14-ethoxymetopon | agonist |
| etonitazene | agonist |
| etorphine | agonist |
| etoxeridine | agonist |
| fentanyl | agonist |
| 4-fluorobutyrfentanyl | agonist |
| 4-fluoropethidine | agonist |
| furanylfentanyl | agonist |
| furethidine | agonist |
| hemorphin-4 | agonist |
| heterocodeine | agonist |
| hydromorphinol | agonist |
| hydromorphone | agonist |
| hydroxypethidine | agonist |
| ibntxa | agonist |
| ic-26 | agonist |
| isomethadone | agonist |
| ketamine | agonist |
| ketobemidone | agonist |
| lefetamine | agonist |
| levacetylmethadol | agonist |
| levallorphan | agonist |
| levomethadone | agonist |
| levophenacylmorphan | agonist |
| levorphanol | agonist |
| lofentanil | agonist |
| loperamide | agonist |
| meprodine | agonist |
| metethoheptazine | agonist |
| methadone | agonist |
| metheptazine | agonist |
| 4-methoxybutyrfentanyl | agonist |
| 14-methoxydihydromorphinone | agonist |
| 14-methoxymetopon | agonist |
| alpha-methylacetylfentanyl | agonist |
| 3-methylbutyrfentanyl | agonist |
| n-methylcarfentanil | agonist |
| methyldesorphine | agonist |
| methyldihydromorphine | agonist |
| 6-methylenedihydrodesoxymorphine | agonist |
| 3-methylfentanyl | agonist |
| beta-methylfentanyl | agonist |
| methylketobemidone | agonist |
| 3-methylthiofentanyl | agonist |
| metopon | agonist |
| mitragynine_pseudoindoxyl | agonist |
| 6-monoacetylmorphine | agonist |
| morpheridine | agonist |
| morphine | agonist |
| morphine-6-glucuronide | agonist |
| morphinone | agonist |
| mr-2096 | agonist |
| mt-45 | agonist |
| noracymethadol | agonist |
| ocfentanil | agonist |
| ohmefentanyl | agonist |
| oliceridine | agonist |
| oxpheneridine | agonist |
| oxymorphazone | agonist |
| oxymorphol | agonist |
| oxymorphone | agonist |
| parafluorofentanyl | agonist |
| pentamorphone | agonist |
| pepap | agonist |
| pethidine | agonist |
| phenadoxone | agonist |
| phenampromide | agonist |
| phenaridine | agonist |
| phenazocine | agonist |
| pheneridine | agonist |
| n-phenethylnordesomorphine | agonist |
| n-phenethylnormorphine | agonist |
| phenomorphan | agonist |
| phenoperidine | agonist |
| 4-phenylfentanyl | agonist |
| 14-phenylpropoxymetopon | agonist |
| picenadol | agonist |
| piminodine | agonist |
| piritramide | agonist |
| prodilidine | agonist |
| prodine | agonist |
| proheptazine | agonist |
| properidine | agonist |
| propylketobemidone | agonist |
| prosidol | agonist |
| pzm21 | agonist |
| r-4066 | agonist |
| r-30490 | agonist |
| racemorphan | agonist |
| remifentanil | agonist |
| ro4-1539 | agonist |
| sc-17599 | agonist |
| semorphone | agonist |
| sufentanil | agonist |
| thienorphine | agonist |
| thiofentanyl | agonist |
| tilidine | agonist |
| trefentanil | agonist |
| trimeperidine | agonist |
| trimu_5 | agonist |
| u-47700 | agonist |
| u-77891 | agonist |
| viminol | agonist |
| 6beta-naltrexol-d4 | antagonist |
| beta-chlornaltrexamine | antagonist |
| beta-funaltrexamine | antagonist |
| alvimopan | antagonist |
| at-076 | antagonist |
| axelopran | antagonist |
| bevenopran | antagonist |
| clocinnamox | antagonist |
| cyclofoxy | antagonist |
| cyprodime | antagonist |
| eptazocine | antagonist |
| ly-255582 | antagonist |
| methocinnamox | antagonist |
| methylnaltrexone | antagonist |
| methylsamidorphan | antagonist |
| nalmefene | antagonist |
| naloxazone | antagonist |
| naloxegol | antagonist |
| naloxol | antagonist |
| naloxonazine | antagonist |
| naloxone | antagonist |
| naltrexazone | antagonist |
| naltrexone | antagonist |
| oxilorphan | antagonist |
| quadazocine | antagonist |
| samidorphan | antagonist |

**Extended Data Table 1**

List of agonists and antagonists used to fit machine learned classifiers. All ligands and labels derived from Wikipedia (https://en.wikipedia.org/wiki/Template:Opioid_receptor_modulators), with some removed on advice by colleagues (cf. Acknowledgements section).

---

a)
*Fentanyl analog ligands (test set):*
['acetylfentanyl', 'acrylfentanyl', '3-allylfentanyl', 'alphamethylthiofentanyl', 'azaprocin', 'betahydroxyfentanyl', 'betahydroxythiofentanyl', 'butyrfentanyl', 'carfentanil', 'desmethylprodine', 'diampromide', 'fentanyl', '4-fluorobutyrfentanyl', 'furanylfentanyl', 'lofentanil', '4-methoxybutyrfentanyl', 'alpha-methylacetylfentanyl', '3-methylbutyrfentanyl', 'n-methylcarfentanil', '3-methylfentanyl', 'beta-methylfentanyl', '3-methylthiofentanyl', 'ocfentanil', 'ohmefentanyl', 'parafluorofentanyl', 'pepap', 'phenampromide', 'phenaridine', '4-phenylfentanyl', 'prodilidine', 'prodine', 'proheptazine', 'prosidol', 'r-30490', 'remifentanil', 'sufentanil', 'thiofentanyl', 'trimeperidine', 'u-47700']

*Non-fentanyl-analog agonists (train set):*
['7-pet', 'alimadol', 'alphamethadol', 'azidomorphine', 'bdpc', 'betamethadol', 'c-8813', 'cebranopadol', 'chloromorphide', 'chloroxymorphamine', 'ciprefadol', 'clonitazene', 'dadle', 'damgo', 'desomorphine', 'dihydroetorphine', 'dihydromorphine', 'dimenoxadol', 'dimepheptanol', 'dimethylaminopivalophenone', 'eluxadoline', 'endomorphin', 'endomorphin-1', '14-ethoxymetopon', 'etonitazene', 'etorphine', 'hemorphin-4', 'heterocodeine', 'hydromorphinol', 'hydromorphone', 'ibntxa', 'ketamine', 'lefetamine', 'levophenacylmorphan', 'levorphanol', '14-methoxydihydromorphinone', '14-methoxymetopon', 'methyldesorphine', 'methyldihydromorphine', '6-methylenedihydrodesoxymorphine', 'metopon', 'mitragynine_pseudoindoxyl', '6-monoacetylmorphine', 'morphine', 'morphine-6-glucuronide', 'morphinone', 'mr-2096', 'oliceridine', 'oxymorphazone', 'oxymorphol', 'oxymorphone', 'pentamorphone', 'phenazocine', 'n-phenethylnordesomorphine', 'n-phenethylnormorphine', 'phenomorphan', '14-phenylpropoxymetopon', 'picenadol', 'pzm21', 'racemorphan', 'ro4-1539', 'sc-17599', 'semorphone', 'thienorphine', 'tilidine', 'trimu_5', 'viminol']

*Antagonists:*
['levallorphan', '6beta-naltrexol-d4', 'beta-chlornaltrexamine', 'beta-funaltrexamine', 'alvimopan', 'at-076', 'axelopran', 'bevenopran', 'clocinnamox', 'cyclofoxy', 'cyprodime', 'eptazocine', 'ly-

255582', 'methocinnamox', 'methylnaltrexone', 'methylsamidorphan', 'nalmefene', 'naloxazone', 'naloxegol', 'naloxol', 'naloxonazine', 'naloxone', 'naltrexazone', 'naltrexone', 'oxilorphan', 'quadazocine', 'samidorphan']

**b)**

*Methadone analog ligands (test set):*
['acetylmethadol', 'alphacetylmethadol', 'alphamethadol', 'betacetylmethadol', 'betamethadol', 'dipipanone', 'ic-26', 'isomethadone', 'ketobemidone', 'levacetylmethadol', 'levomethadone', 'methadone', 'methylketobemidone', 'noracymethadol', 'phenadoxone', 'propylketobemidone', 'r-4066']

*Non-methadone analogs (train set):*
['7-pet', 'alimadol', 'azidomorphine', 'bdpc', 'c-8813', 'cebranopadol', 'chloromorphide', 'chloroxymorphamine', 'ciprefadol', 'clonitazene', 'dadle', 'damgo', 'desomorphine', 'dihydroetorphine', 'dihydromorphine', 'dimenoxadol', 'dimepheptanol', 'dimethylaminopivalophenone', 'eluxadoline', 'endomorphin', 'endomorphin-1', '14-ethoxymetopon', 'etonitazene', 'etorphine', 'hemorphin-4', 'heterocodeine', 'hydromorphinol', 'hydromorphone', 'ibntxa', 'ketamine', 'lefetamine', 'levophenacylmorphan', 'levorphanol', '14-methoxydihydromorphinone', '14-methoxymetopon', 'methyldesorphine', 'methyldihydromorphine', '6-methylenedihydrodesoxymorphine', 'metopon', 'mitragynine_pseudoindoxyl', '6-monoacetylmorphine', 'morphine', 'morphine-6-glucuronide', 'morphinone', 'mr-2096', 'oliceridine', 'oxymorphazone', 'oxymorphol', 'oxymorphone', 'pentamorphone', 'phenazocine', 'n-phenethylnordesomorphine', 'n-phenethylnormorphine', 'phenomorphan', '14-phenylpropoxymetopon', 'picenadol', 'pzm21', 'racemorphan', 'ro4-1539', 'sc-17599', 'semorphone', 'thienorphine', 'tilidine', 'trimu_5', 'viminol']

Antagonists:
['levallorphan', '6beta-naltrexol-d4', 'beta-chlornaltrexamine', 'beta-funaltrexamine', 'alvimopan', 'at-076', 'axelopran', 'bevenopran', 'clocinnamox', 'cyclofoxy', 'cyprodime', 'eptazocine', 'ly-255582', 'methocinnamox', 'methylnaltrexone', 'methylsamidorphan', 'nalmefene', 'naloxazone', 'naloxegol', 'naloxol', 'naloxonazine', 'naloxone', 'naltrexazone', 'naltrexone', 'oxilorphan', 'quadazocine', 'samidorphan']

**Extended Data Table 2**

**a)**

A scaffold split was defined in which (1) agonist ligands with a Tanimoto score of ≤0.5 compared to fentanyl were placed in a train set, (2) agonist ligands with a Tanimoto score of ≥0.7 compared to fentanyl were placed in a test set, and (3) antagonists were randomly distributed between the train and test sets.

**b)** A scaffold split was defined in which (1) agonist ligands with a Tanimoto score of ≤0.5 compared to methadone were placed in a train set, (2) agonist ligands with a Tanimoto score of

≥0.7 compared to methadone were placed in a test set, and (3) antagonists were randomly distributed between the train and test sets.

a)

| | |
|---|---|
| Inactive Crystal | 0.063488 |
| State 14 | 0.033463 |
| State 3 | 0.031175 |
| State 17 | 0.029950 |
| State 10 | 0.029853 |
| State 23 | 0.025154 |
| State 5 | 0.024361 |
| State 16 | 0.023912 |
| State 21 | 0.023884 |
| State 4 | 0.021384 |
| State 22 | 0.020618 |
| State 0 | 0.019934 |
| State 13 | 0.019720 |
| State 18 | 0.017975 |
| State 7 | 0.017955 |
| State 24 | 0.017434 |
| State 11 | 0.017295 |
| State 9 | 0.016170 |
| State 8 | 0.015486 |
| State 15 | 0.015193 |
| State 19 | 0.013673 |
| Active Crystal | 0.013580 |
| State 12 | 0.013346 |
| State 6 | 0.012534 |
| State 2 | 0.012306 |
| State 1 | 0.012289 |
| State 20 | 0.011230 |

b)

| | |
|---|---|
| Inactive Crystal | 0.057546 |
| State 18 | 0.051126 |
| State 2 | 0.050497 |
| State 14 | 0.045177 |
| State 1 | 0.043236 |
| Active Crystal | 0.040393 |
| State 20 | 0.039999 |
| State 3 | 0.037063 |
| State 24 | 0.035946 |
| State 19 | 0.035883 |
| State 17 | 0.035723 |
| State 5 | 0.035547 |
| State 22 | 0.034922 |
| State 16 | 0.034121 |
| State 0 | 0.034035 |
| State 21 | 0.033999 |
| State 4 | 0.033884 |
| State 6 | 0.033837 |
| State 13 | 0.033381 |

```
State 10               0.032723
State 11               0.032537
State 23               0.032248
State 12               0.032069
State 9                0.031896
State 15               0.031409
State 8                0.030558
State 7                0.030245
```

**Extended Data Table 3**

Random Forest average Gini impurity reduction ("importance") of each feature (MD State, Crystal Structure) for **a)** distinguishing between opioid agonists and antagonists and **b)** distinguishing between binders and non-binders from µOR.

| Dataset | Cross validation split type | AUC (Crystals alone) | AUC (Crystal + MD structures) | Wilson 99% CI |
|---|---|---|---|---|
| Wikipedia Agonists/Antagonists | Random | 0.72 | 0.86 | (0.82, 0.88) |
| Wikipedia Agonists/Antagonists | Methadone | 0.84 | 0.99 | (0.82, 0.88) |
| Wikipedia Agonists/Antagonists | Fentanyl | 0.77 | 0.93 | (0.98, 1.0) |
| Expert curated dataset | Random | 0.73 | 0.85 | (0.67, 0.75) |
| Expert curated dataset | Methadone | 0.89 | 0.94 | (0.51, 0.59) |
| Expert curated dataset | Fentanyl | 0.81 | 0.91 | (0.88, 0.93) |

**Extended Data Table 4**

Docking to both MD states and crystal structures statistically significantly improves ability over crystals alone to distinguish µOR binders from non-binders. Table shows median ROC Area Under the Curve (AUC) performance over the validation set over 1,000 train-valid splits for different split and model types. Differences between crystal lone and crystal + MD structures methods are considered statistically significant if the lower bound of a 99% Wilson scoring confidence interval (CI) is greater than 0.5. Note that, for each dataset, incorporating MD-derived structures in addition to the crystal structures confers a statistically significant improvement in ability to distinguish binders from non-binder as measured by AUC. Notably, when fentanyl (or methadone) analogs are removed from the training set, models remain able to distinguish fentanyl (or methadone) derivative agonists from random sets of antagonists. This indicates that models fit in this way have the capacity to discover new opioid agonist scaffolds in addition to derivatives of existing ones.

| Dataset | Split | AUC (Crystals alone) | AUC (Crystal + MD structures) | Wilson 99% CI |
|---|---|---|---|---|
| Measured pIC50, cutoff = 6.0 | Random | 0.59 | 0.64 | (0.76, 0.82) |
| Measured pIC50, cutoff = 7.0 | Random | 0.59 | 0.71 | (0.99, 1.0) |
| Measured pIC50 cutoff = 8.0 | Random | 0.58 | 0.74 | (0.99, 1.0) |
| All, pIC50 cutoff = 5.3 | Random | 0.78 | 0.87 | (0.99, 1.0) |
| All, pIC50 cutoff = 6.0 | Random | 0.73 | 0.82 | (0.99, 1.0) |
| All, pIC50 cutoff = 7.0 | Random | 0.67 | 0.78 | (0.99, 1.0) |
| All, pIC50 cutoff = 8.0 | Random | 0.65 | 0.79 | (0.99, 1.0) |
| All, pIC50 cutoff = 5.3 | Scaffold | 0.77 | 0.81 | (0.73, 0.80) |
| All, pIC50 cutoff = 6.0 | Scaffold | 0.78 | 0.83 | (0.73, 0.8) |
| All, pIC50 cutoff = 7.0 | Scaffold | 0.66 | 0.79 | (0.86, 0.91) |
| All, pIC50 cutoff = 8.0 | Scaffold | 0.64 | 0.78 | (0.85, 0.9) |

**Extended Data Table 5**

Docking to both MD states and crystal structures statistically significantly improves ability over crystals alone to distinguish μOR binders from non-binders. Table shows median ROC Area Under the Curve (AUC) performance over the validation set over 1,000 train-valid splits for different split and model types. Differences between crystal lone and crystal + MD structures methods are considered statistically significant if the lower bound of a 99% Wilson scoring confidence interval (CI) is greater than 0.5. Note that, for each dataset, incorporating MD-derived structures in addition to the crystal structures confers a statistically significant improvement in ability to distinguish binders from non-binder as measured by AUC. Notably, when molecules with similar scaffolds (as measured by a Tanimoto similarity score of > 0.7) are removed from the training data, models remain able to distinguish binders from non-binders. This indicates that models fit in this way have the capacity to discover new opioid scaffolds in addition to derivatives of existing ones.

Datasets consist of compounds with experimentally known values of binding affinity to μOR according to Chembl, Wikipedia, and other publicly available sources (full list of ligands in this dataset available in supplemental files). Datasets termed "Measured Ki" include only those compounds with a real-numbered Ki value listed in Chembl; datasets termed "All" also include compounds that have no listed Ki but are termed "Not Active" by Chembl. Therefore, "Measured Ki" datasets are subsets of the "All" series of datasets. Binders are considered to be compounds with a pIC50 greater than some cutoff (listed in the "Dataset" table) and non-binders with a pIC50 lower than that same cutoff. For example, "All, pIC50 cutoff = 7.0" indicates a dataset wherein (a) both ligands with a measurable pIC50 < 7.0 and those listed as "Not Active" in Chembl are considered non-binders, and (b) both ligands with a measurable pIC50 ≥ 7.0 and other known agonists and antagonists from Wikipedia are considered to be binders.

| FMP # | Supplier | Catalogue no. | Molecular Weight | Structure |
|---|---|---|---|---|
| 1 | ChemBridge Corp | 5315117 | 261.409 | 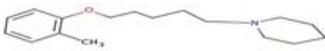 |

| 2 | Vitas-M Laboratory | STK246443 | 220.272 | 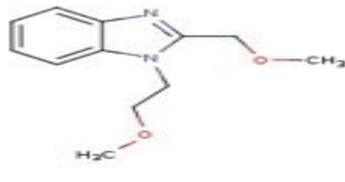 |
|---|---|---|---|---|
| 3 | ChemBridge Corp | 5421855 | 298.471 | 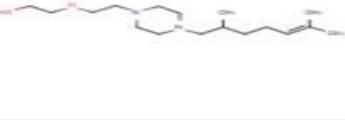 |
| 4 | ChemBridge Corp | 5359784 | 286.419 | 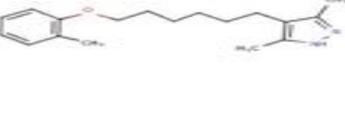 |
| 5 | ChemDiv, Inc. | K416-0034 | 331.45 | 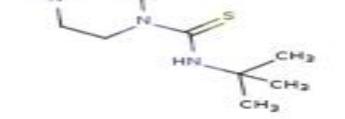 |
| 6 | ChemDiv, Inc. | 6030-4029 | 277.408 | 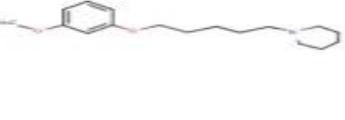 |
| 7 | ChemDiv, Inc. | D171-0103 | 284.403 | 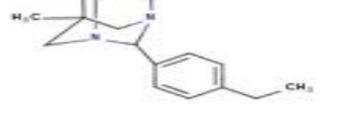 |
| 8 | ChemDiv, Inc. | K808-9296 | 317.388 | 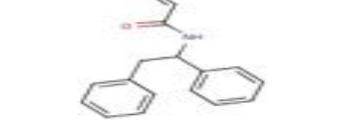 |

| 9 | ChemDiv, Inc. | C296-0706 | 379.43 | 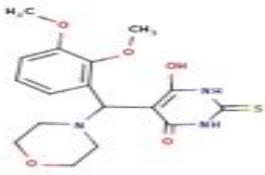 |
| --- | --- | --- | --- | --- |
| 10 | ChemDiv, Inc. | C656-0381 | 347.353 | 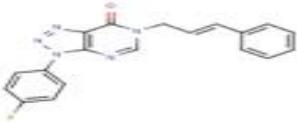 |
| 11 | InterBioScreen Ltd. | STOCK1S-36272 | 255.4 | 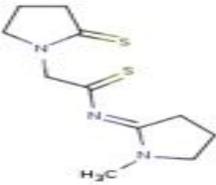 |
| 12 | Life Chemicals Inc. | F5959-0041 | 249.33 | 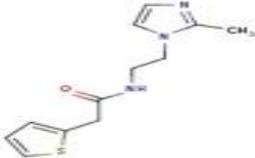 |
| 13 | Life Chemicals Inc. | F6350-0021 | 228.31 | 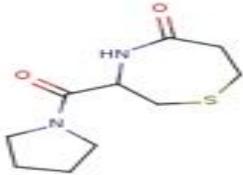 |
| 14 | MayBridge, Ltd. | S14427 | 325.324 | 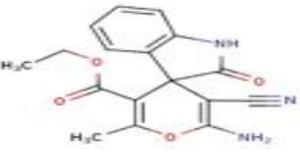 |
| 15 | MayBridge, Ltd. | CC52213 | 207.273 | 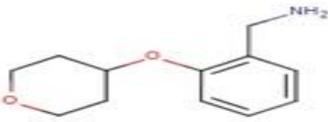 |

| 16 | Specs | AG-644/14117620 | 286.419 | 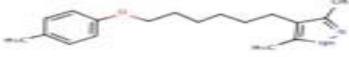 |
| --- | --- | --- | --- | --- |
| 17 | Specs | AF-399/41900457 | 282.77 | 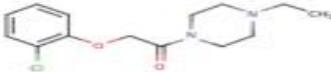 |
| 18 | Specs | AK-968/13031275 | 335.23 | 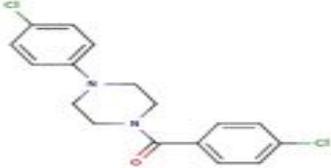 |
| 19 | Specs | AG-690/11549877 | 265.353 | 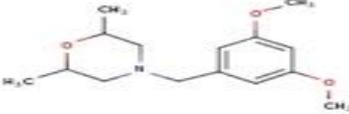 |
| 20 | Specs | AF-399/41875794 | 291.435 | 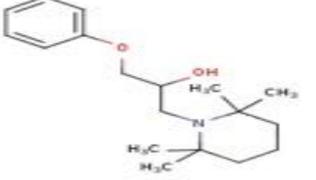 |
| 21 | Specs | AG-664/15584112 | 290.451 | 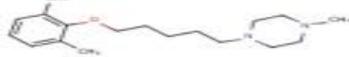 |
| 22 | Specs | AG-690/11764068 | 308.422 | 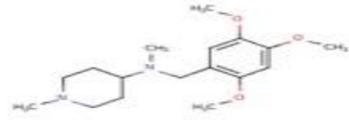 |

| 23 | Vitas-M Laboratory | STK768975 | 354.47 | 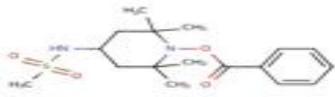 |
| 24 | Vitas-M Laboratory | STK825555 | 408.28 | 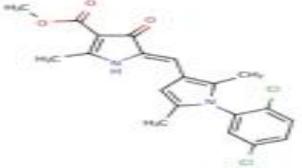 |
| 25 | Vitas-M Laboratory | STK705512 | 390.399 | 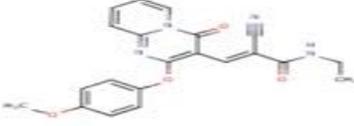 |
| 26 | Vitas-M Laboratory | STL181788 | 309.453 | 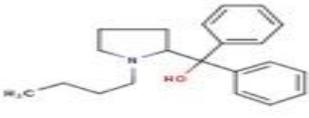 |
| 27 | Vitas-M Laboratory | STK123054 | 253.301 | 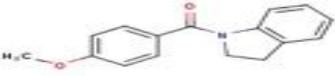 |
| 28 | Vitas-M Laboratory | STK386756 | 343.28 | 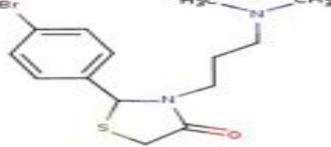 |
| 29 | Vitas-M Laboratory | STK766152 | 285.387 | 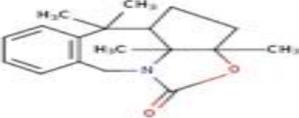 |

| 30 | ChemBridge Corp | 5316199 | 263.381 | 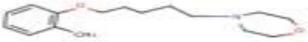 |
| 31 | ChemBridge Corp | 5455149 | 225 | 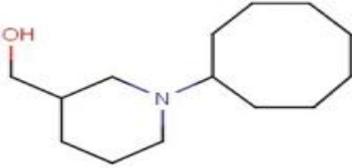 |
| 32 | ChemBridge Corp | 5410648 | 280 | 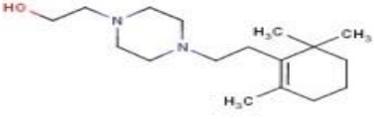 |

**Extended Data Table 6**

List of all 32 purchased compounds that were experimentally tested for binding.

|  |  |  | Binding Data - $K_i$ (nM)[a] | | |
| --- | --- | --- | --- | --- | --- |
| FMP# | Structure | MW | MOR-1 | KOR-1 | DOR-1 |
| 1 | 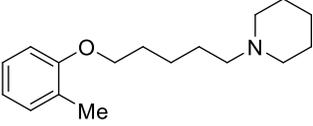 | 261.409 | 841.4 ±241.9 | 7386 ±2376 | 17874 ±7615 |
| 4 | 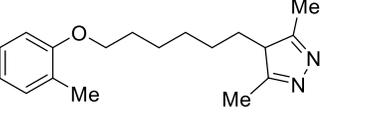 | 286.419 | 3217 ±153 | 2503 ±523 | 8143 ±1398 |
| 6 | 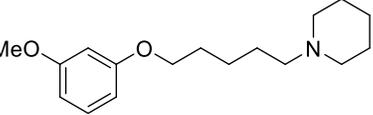 | 277.408 | >30 µM | 6941 ±1970 | >30 µM |
| 16 | 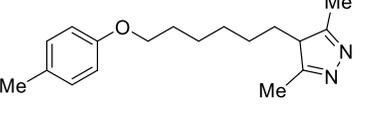 | 286.419 | 1748 ±492 | 4918 ±2235 | 9058 ±566 |
| 21 | 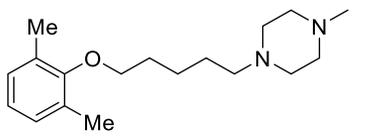 | 290.451 | >30 µM | 5503 ±4194 | >30 µM |
| 30 | 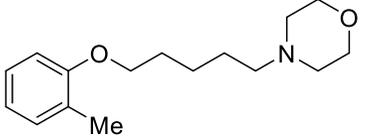 | 263.381 | 11315 ±2262 | 6009 ±2163 | 12761 ±856 |
| | DAMGO | | 3.3 ± 0.43 [b] | | |
| | U50,488h | | | 0.73 ± 0.32 [b] | |
| | DPDPE | | | | 1.39 ± 0.67 [b] |

**Extended Table Data 7. Receptor binding of FMP4 and structurally similar analogs**

[a] Competition studies were performed with the indicated compounds against [125]I-IBNtxA (0.1 nM) in membranes from CHO cells stably expressing the indicated cloned mouse opioid receptors. Results are presented as nM ± SEM from three independent experiments performed in triplicate. [b] Values from the literature.[8]